RUSSIAN ACADEMY OF SCIENCES
PETERSBURG NUCLEAR PHYSICS INSTITUTE

G. I. Shulyak, A. A. Rodionov

# The ENSDF_toolbox program package: tool for the evaluator of nuclear data

**A b s t r a c t**

The program package for the work with the Evaluated Nuclear Structure Data File is discussed. The program shell designed for the unification of the process of the evaluation of the nuclear data is proposed. This program shell may be used in the regular work of the nuclear data evaluator and for common use by scientists and engineers who need the actual data about nuclear states and transitions from the ENSDF database.

## Introduction

The wide usage of achievements of nuclear physics in basic researches, technology, medicine and other activities supposes the access for specialists in these fields of knowledge to the information which is collected in the nuclear databases. Now, nuclear physics has accumulated a huge experimental material. Parameters of the ground and excited states of atomic nuclei, the cross sections of interaction of subatomic particles with nuclei and other values determined in experiments with atomic nuclei are published in scientific journals and books. Various experiments use different devices and methods, but investigate different properties of the same objects, namely, atomic nuclei.

Thus, there is a problem of obtaining from the published data the most reliable, complete and consistent picture of the nuclear states and reactions with nuclei. More than fifty years ago the work on evaluation of the nuclear data was begun by scientists from some groups in the USA. The result of the work of the evaluator, who should be a qualified physicist, is an adopted scheme of excited states of nucleus and adopted values of other parameters of atomic nuclei. The evaluation is based on all the publications about the given nucleus.

Increasing of the amount of experimental material, sophistication and standardization of the process of evaluation had the result in the expansion of the list of evaluators and usage of computers for processing and storage of the information. Under the auspices of the IAEA (details can be found at the site http://www-nds.iaea.org) the International Network of Nuclear Structure and Decay Data Evaluators was created. The results of efforts of all the evaluators are summarized in the file of evaluated data on the structure of the nuclei (Evaluated Nuclear Structure Data File - ENSDF) and the file ENDF, which contains data on nuclear reactions. Files are supported by the National Nuclear Data Center, Brookhaven National Laboratory (USA). These files are scientific databases with the open access for reading through the Internet. Files are supported in actual conditions. Information is renewed with a period of few years, or if new information appears. The journal *Nuclear Data Sheets* is based on the ENSDF database. It is the well known result of work of the International Network of Nuclear Structure and Decay Data Evaluators.

## The structure of the ENSDF and processing of the nuclear data

The Evaluated Nuclear Structure Data File is a text file which consists of 80-symbol strings named “record” or “card”. The length of a record equal to 80 symbols comes to an ancient time when that file had been created and saved as an array of pasteboard cards with a perforation. In this paper both terms are equivalent. In general case the file contains next datasets, see [1]:

1. General information for a given mass chain, e.g., the name and affiliation of the evaluator, cutoff date, evaluator's remarks, details of the publication, etc.

2. The references used in all the data sets for the given mass number. This data set is based upon reference codes (keynumbers). These codes are the result of the work on the referring of the publications in nuclear physics (NSR; Nuclear Science References); the project is guided by NNDC BNL. The list of refs which are used in various data sets for a given mass number is added to the file by NNDC on the basis of the NSR database.

3. Adopted data sets giving adopted properties of the levels and radiations for the given nucleus.

4. The evaluated results of a single experiment, e.g., a radioactive decay or a nuclear reaction.

5. The combined evaluated results of a number of experiments of the same kind, e.g., reactions with heavy ions, Coulomb excitation, etc., for the given nucleus.

The full description of the ENSDF, data formats and examples are given in [1]. The general organization of the ENSDF is shown schematically in fig. 1.

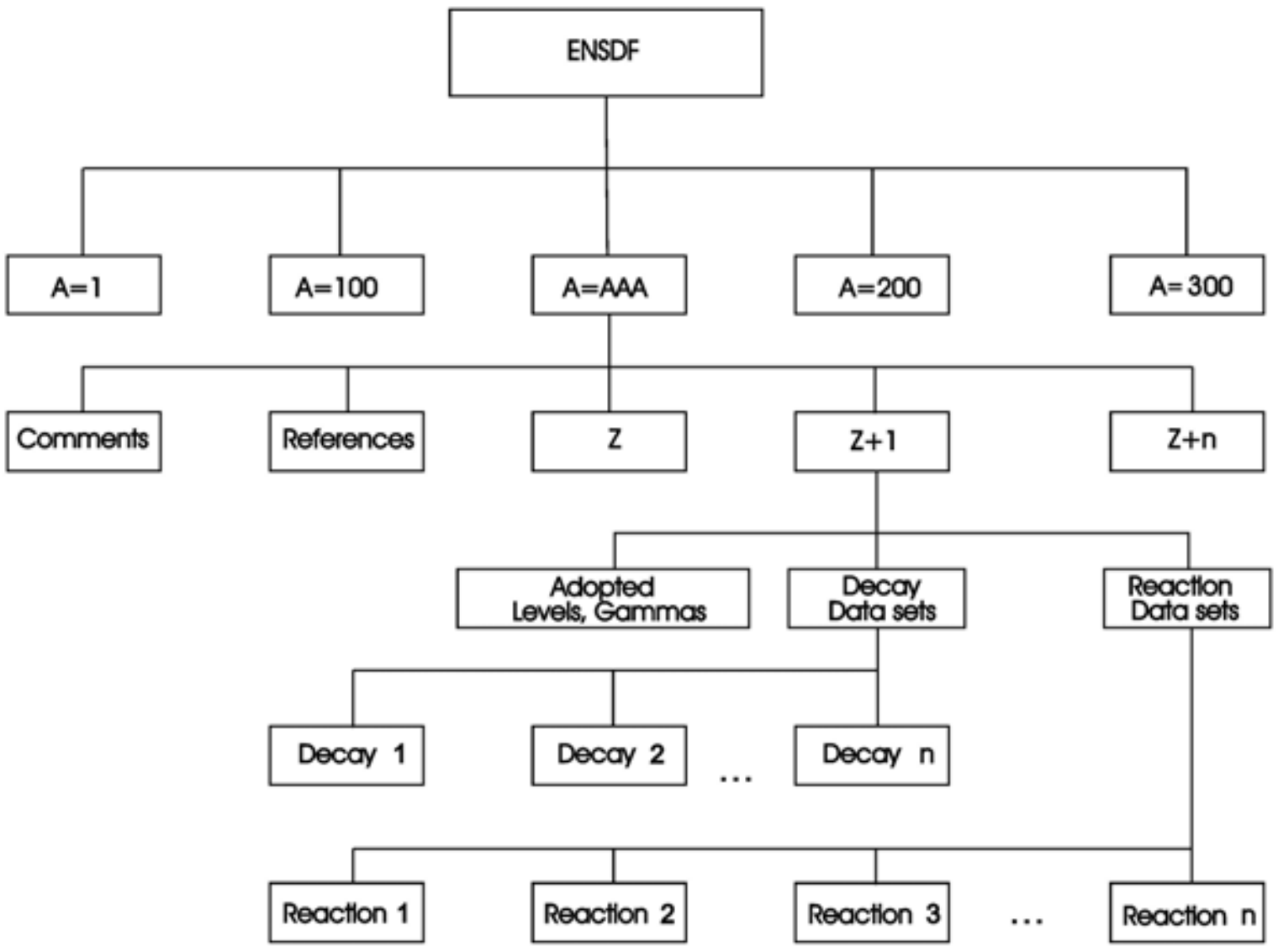

Fig. 1

Process of the evaluation includes the analysis of old and new experimental data from the point of view of the physical significance and the editing of the data sets in accordance with rules defined in [1]. During the work the evaluator uses different programs for checking the dataset for syntax, for calculations of the level energies through the experimental values of energies of transitions, of the intensity balance, of values log *ft*, for the determination and evaluation of the reliability of other parameters, and so on. At last, the evaluator has the file in the PS or PDF format and this file presents the (approximate) copy of the corresponding pages of the *Nuclear Data Sheets* journal. Programs for the data processing may be found at the site of BNL http://www.nndc.bnl.gov. Here these programs are called Standard BNL programs. Besides, the evaluator uses his own programs to solve his problems. These programs may be included afterwards into the list of Standard programs and presented at the BNL site for the common use. Programs are created for different operating systems. Most used systems are MS Windows and Linux. All these programs may be classified as programs for the data analysis and the utility programs which are commonly used for the preparing of the dataset. Here Standard programs are listed with a short description:

**Programs for the data analysis:**

ALPHAD  Calculates hindrance factors for α-decay and the theoretical value of $T_{1/2}$; determines $r_0$ for transitions between even-even ground states using expressions from [2].

HSICC  Calculates coefficients of the internal conversion by the method [3]; now is replaced by the BrIcc program.

BrIcc  Calculates coefficients of the internal conversion [4].

LOGFT  Calculates the values of log *ft* for β-transitions and average energies for the β-radiation [5].

DELTA  Analyses angular correlations and mixing ratios; gives the best values for the mixing ratios.

PANDORA  Checks inconsistency of the data.

GABS  Calculates absolute intensities of γ-rays and the normalization factor for the transferring of the relative intensities to the intensities per 100 decays of the parent nucleus.

RadList  Calculates the parameters of radiation caused by the radioactive decay of the nucleus (energies, intensities, dose rates).

GTOL  Calculates the level energies from the measured values of energies of transitions, determines the intensity balance.

RULER  Calculates the reduced strength for the electromagnetic transitions.

**Utilities:**

ADDGAM  Adds the transitions to the Adopted Levels dataset from one or some datasets.

FMTCHK  Checks the ENSDF syntax of the file; does some other checks.

COMTRANS Translates the special ENSDF format file (see [1]) to the rich text type format file.

TREND Prints tables from the ENSDF, does not makes the PostScript output.

ENSDAT Translates the ENSDF format file to the PostScript file. This file looks as pages in the *Nuclear Data Sheets* journal.

The flowchart in the fig. 2 presents the order of the usage the Standard program package when the Adopted Levels dataset is to be prepared or edited. Similar flowcharts illustrate the processing of the data for the decay and reactions datasets [6].

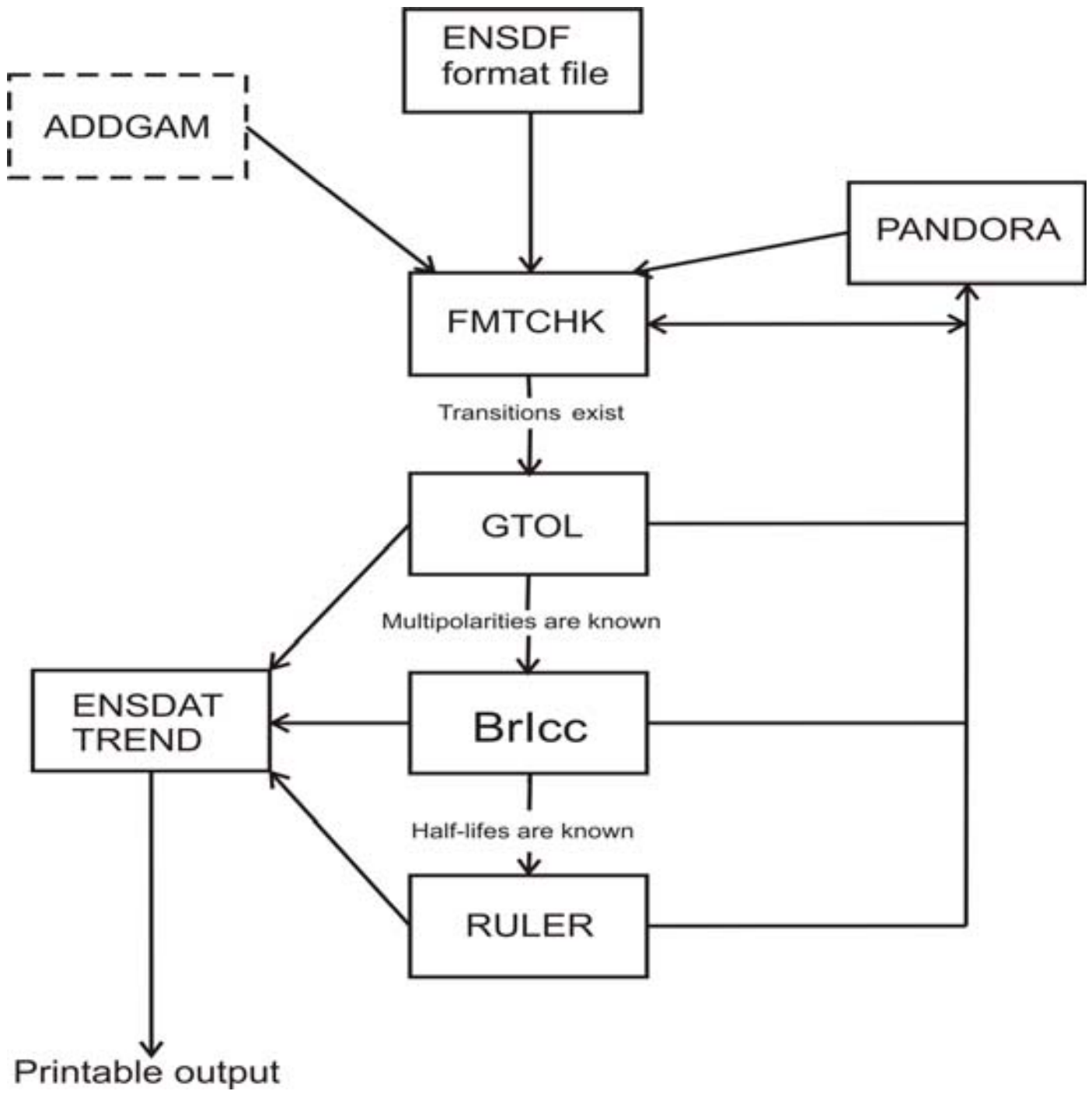

Fig. 2

All the programs listed above had been developed for the work from the system console. After the start of the program the necessary parameters and file names should be typed in the terminal dialog and the result of calculations will be written in the text files. These files may be read in any text viewer or editor. Usually the same editor is used for the editing of the ENSDF format file.

Such an approach was the only one possible 20 or 30 years ago but now it should not be accepted as satisfactory and comfortable. A variety of programs, huge amount of filenames and parameters which the evaluator should retain makes the work hard and leads to the appearance of misprints in the resulting file. It should be marked also that the Standard programs and ENSDF database are widely used by physicists and other specialists for the building and analysis of the level schemes and for obtaining of the necessary nuclear parameters. The evident way is the integration of the Standard package and data from the ENSDF into the unified package as it is made for designer and office workplaces.

The program package proposed in this paper simplifies the process of evaluation of nuclear data, diminishes the number of misprints in the ENSDF, and, in addition, makes easy the access to the nuclear data for any user who needs that information.

## The ENSDF_toolbox program

The ENSDF_toolbox program package is written in C language and operates in the Linux operation system, any graphical environment with GNOME/GTK+ libraries. The main window of the ENSDF_toolbox program looks like the panel with buttons and pull-down menus (fig. 3). It is possible from the main panel and subordinate panels to realize the process of evaluation in the order which is determined by the flowchart fig. 2 or in any other order needed. The user may:

a) to create new datasets, to load datasets from the ENSDF and other datasets of the ENSDF format;

b) to select the data as whole datasets and as defined records in the dataset;

c) to check and to process the data with external programs such as the Standard BNL programs and special ENSDF editor included into the package;

d) to save datasets and results of the external programs.

Here the term “external program” means the program which may be included into the program package if the proper description is made in the proper configuration text file by the user or somebody else. Rules for the introduction of the program into the package may be found in the downloaded package in the file doc/en/panels_description.

Terms “Initial Data” and “Selected Data” are used in the ENSDF_toolbox program. These data are the datasets in the ENSDF format. The program gives the unified way for selection of the data by the request of the user. For this purpose the file “Initial Data” is loaded and the operation of the selection is applied. The result of the operation will be “Selected Data”. The file “Initial Data” is always the initial file for the selection process. If it is necessary to repeat the selection upon the selected information (that is to apply consequent filters), the menu item “Tool/Selected to Initial” should be used. The real selection of the data is initialized by the command “Apply” at the main panel. The selected data may be saved in the user defined file.

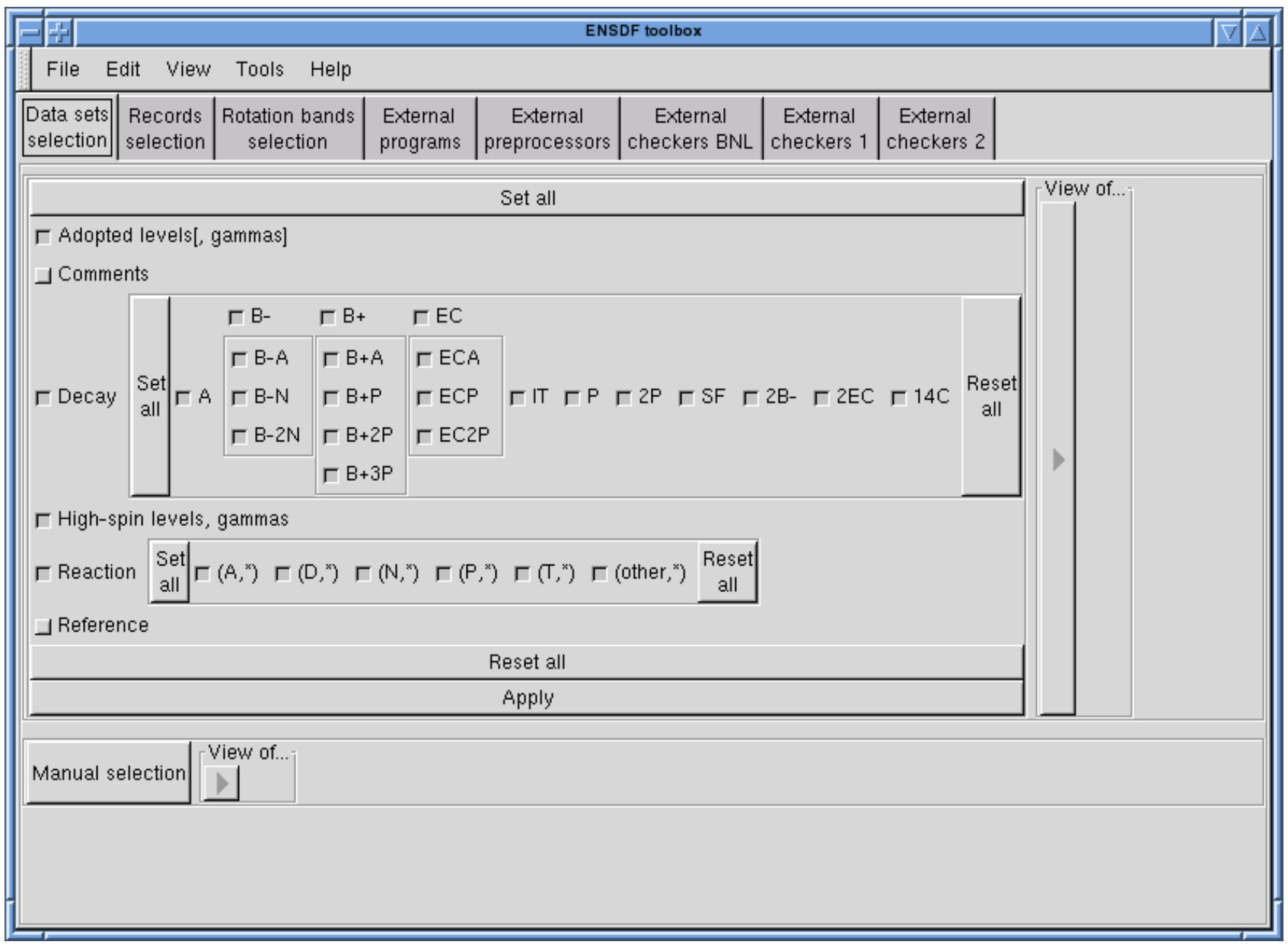

Fig. 3

Figures 4, 5 and 6 present panels for the calling of the programs for the analysis of the data and utilities (Standard BNL programs) and of some other programs. The example of the PostScript output is shown in fig. 7. It is the output file of the ENSDAT program which may be called by the corresponding button on the panel fig. 5.

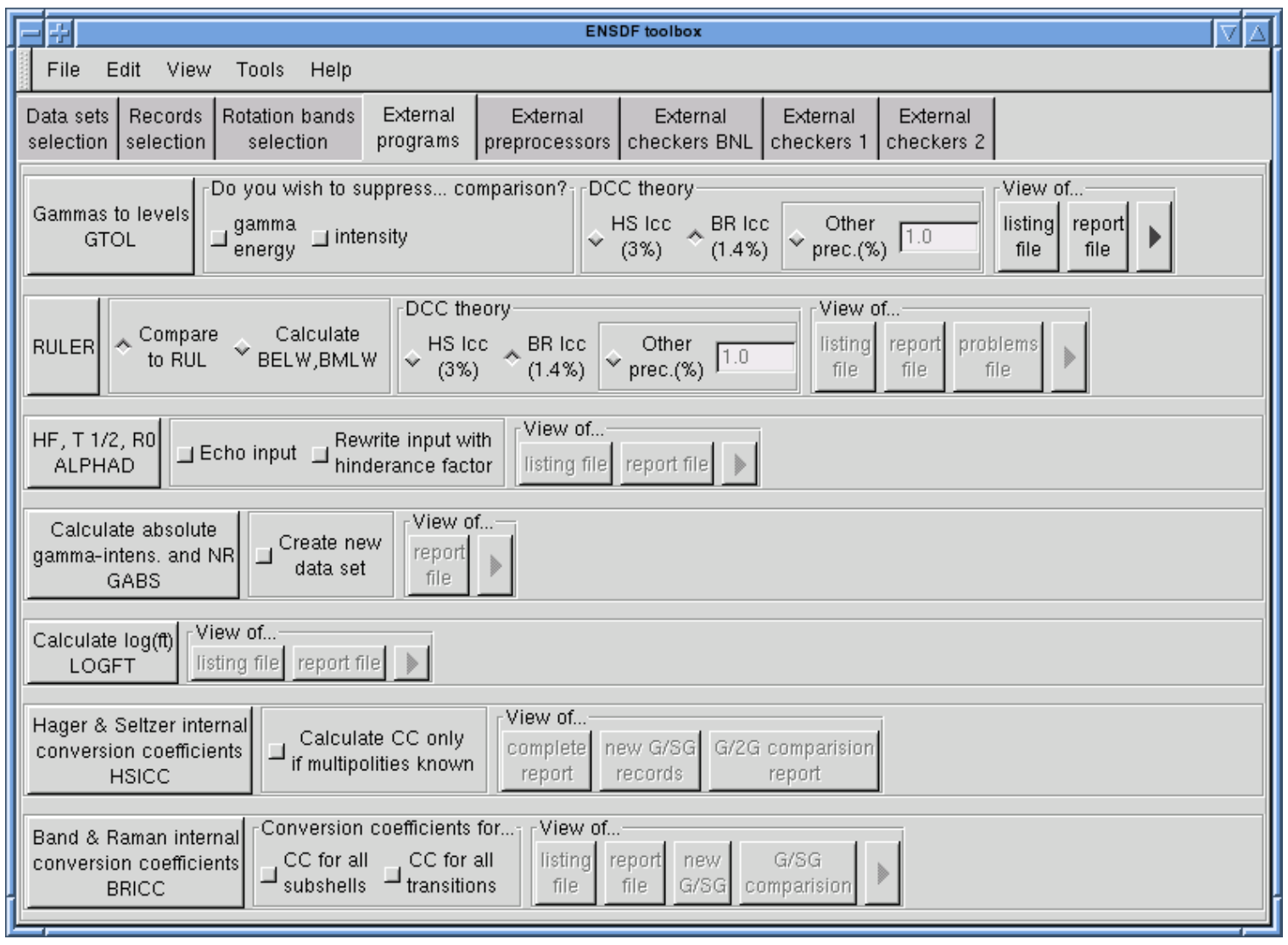

Fig. 4

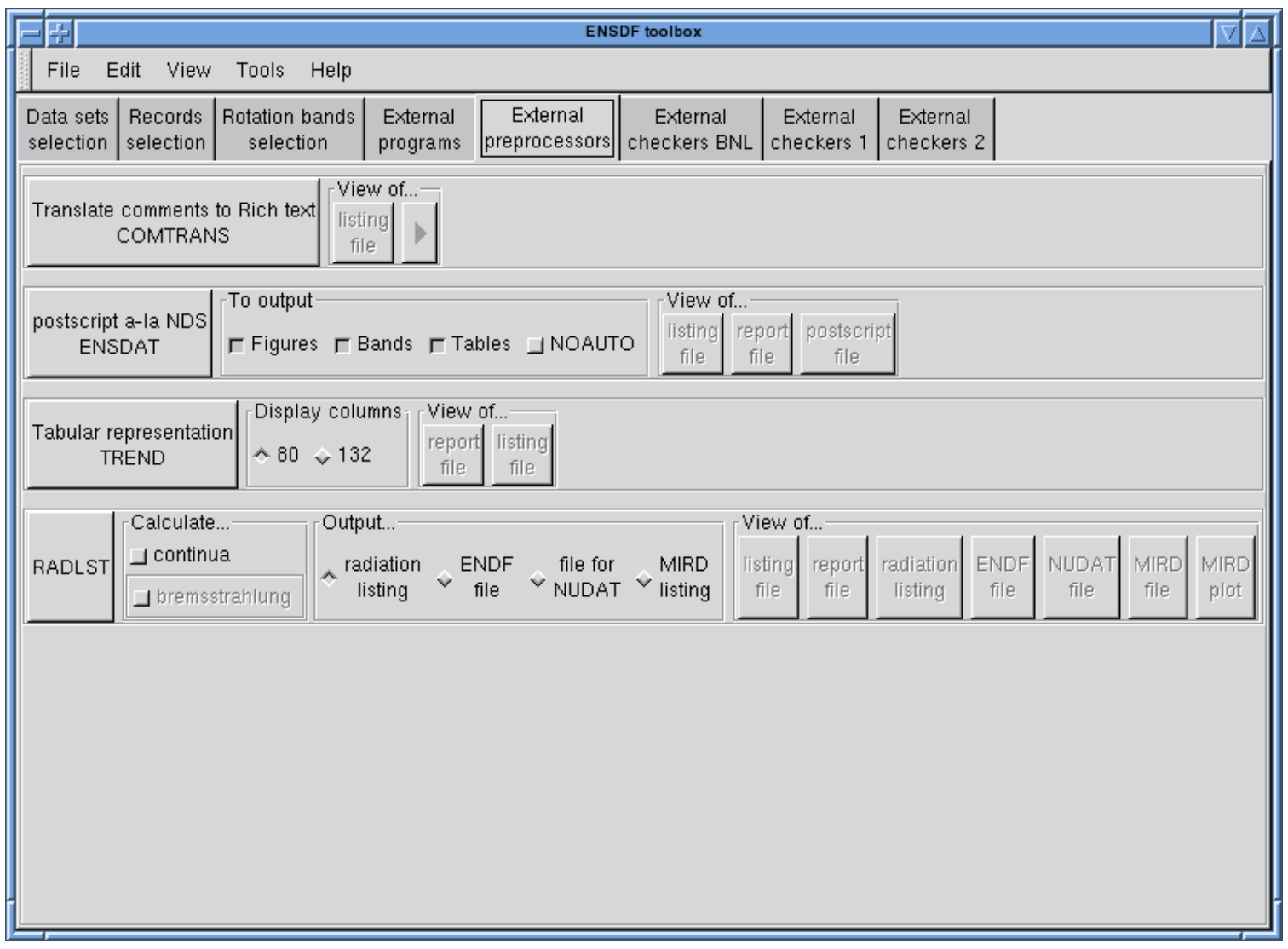

Fig. 5

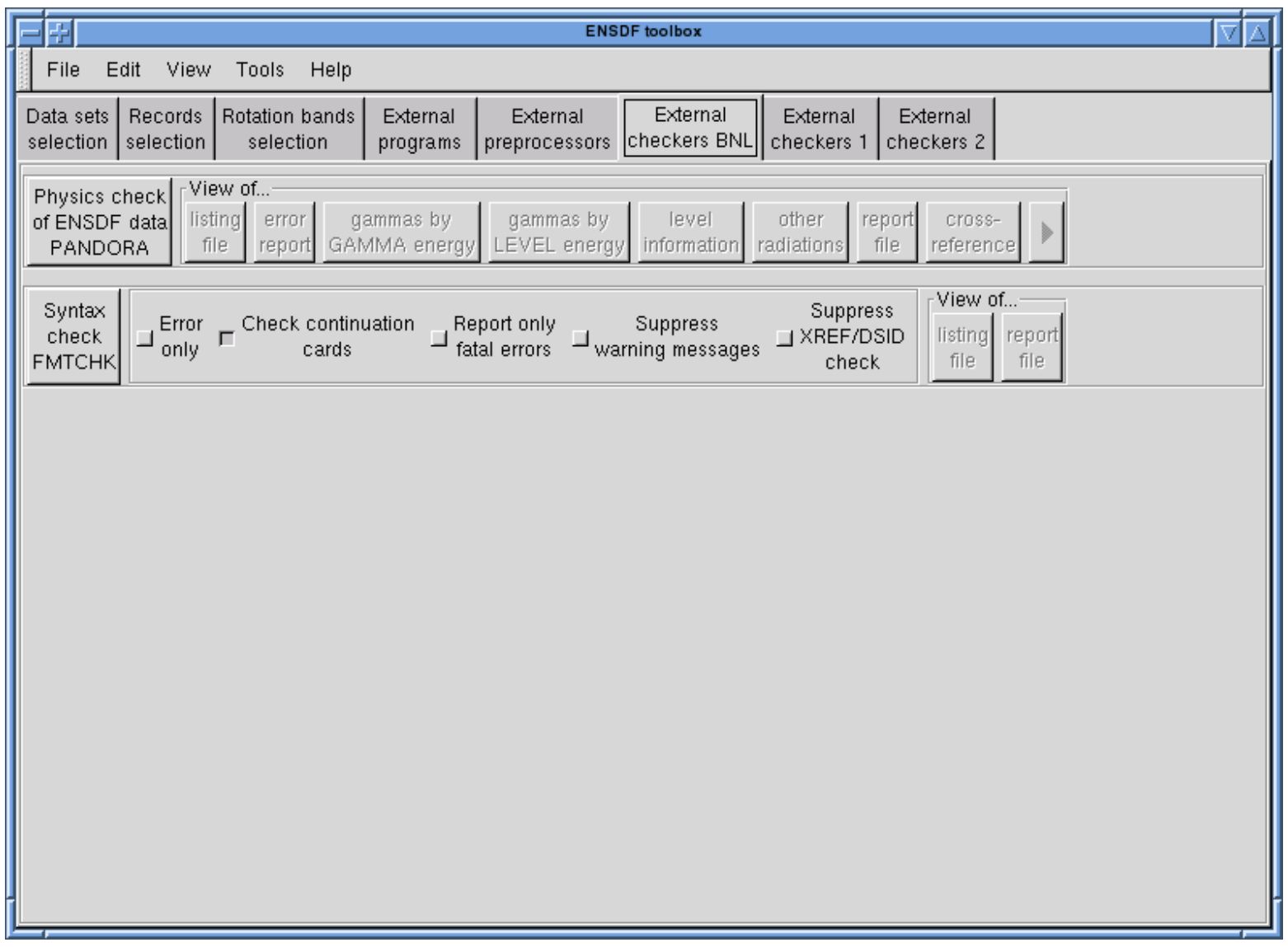

Fig. 6

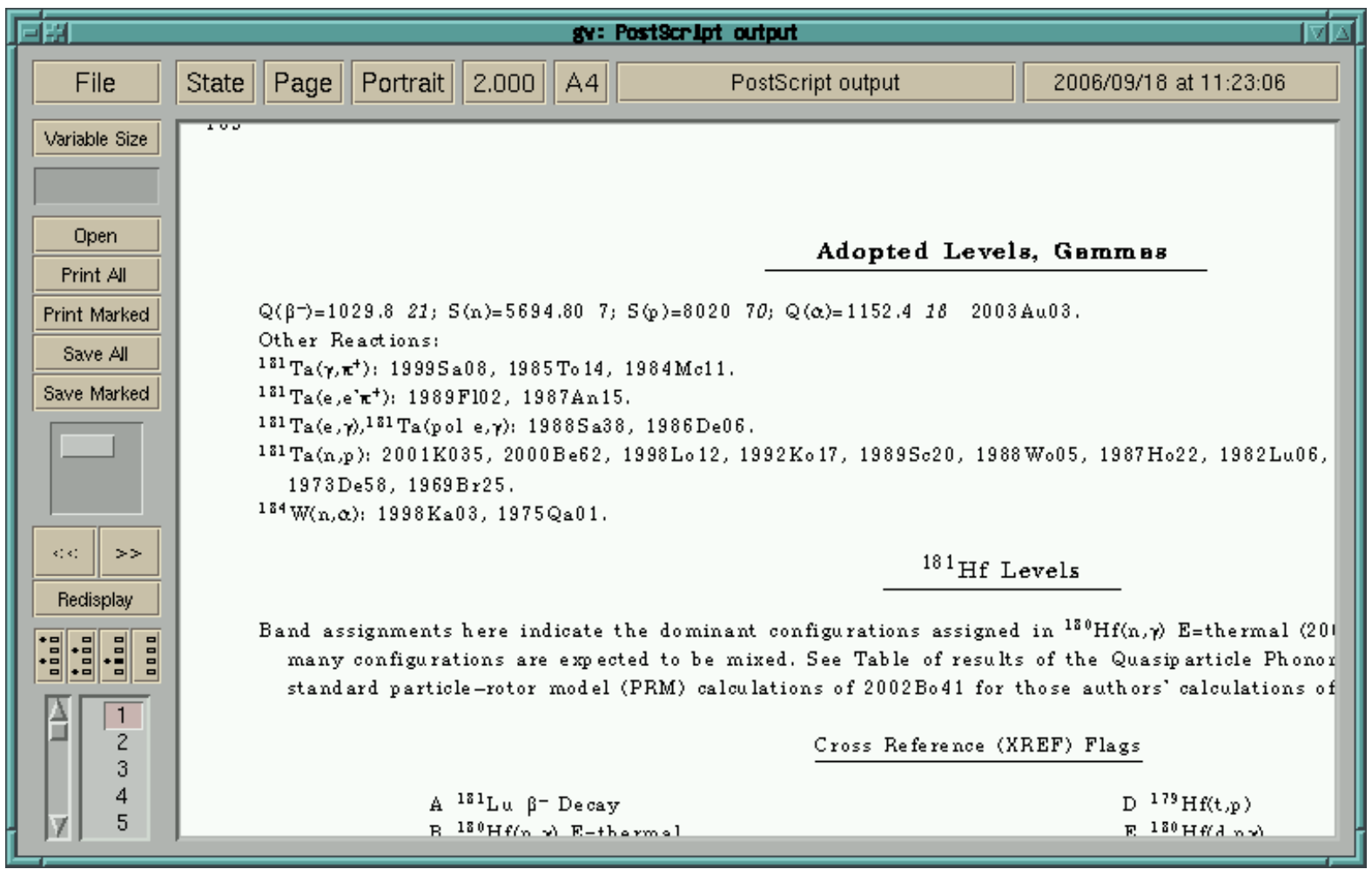

Fig. 7

Among external programs we would like to note the ENSDF editor EVE and the program data_set_viewer [7]. These specialized programs themselves may be interesting for evaluators and may be used apart from the ENSDF_toolbox program package.

The program EVE is the ENSDF formatted text editor. The main difference of EVE from the general purpose text editors is the organization of the working place. Fields and lines of the editor are context-dependent which allow checking the file to the ENSDF syntax “on the fly”. The EVE editor is completely tunable by the user if necessary. The rules for checking, colors of fields and other parameters are determined in the configuration text file and may be easily changed or completed, if the syntax is changed or, if the new card should be described.

The program data_set_viewer is the level scheme viewer. It builds the level scheme from the ENSDF file. The picture may be shifted or zoomed if the detail view is desirable, see fig. 8. This program may be called from the menu “View” at panels figs. 3-6.

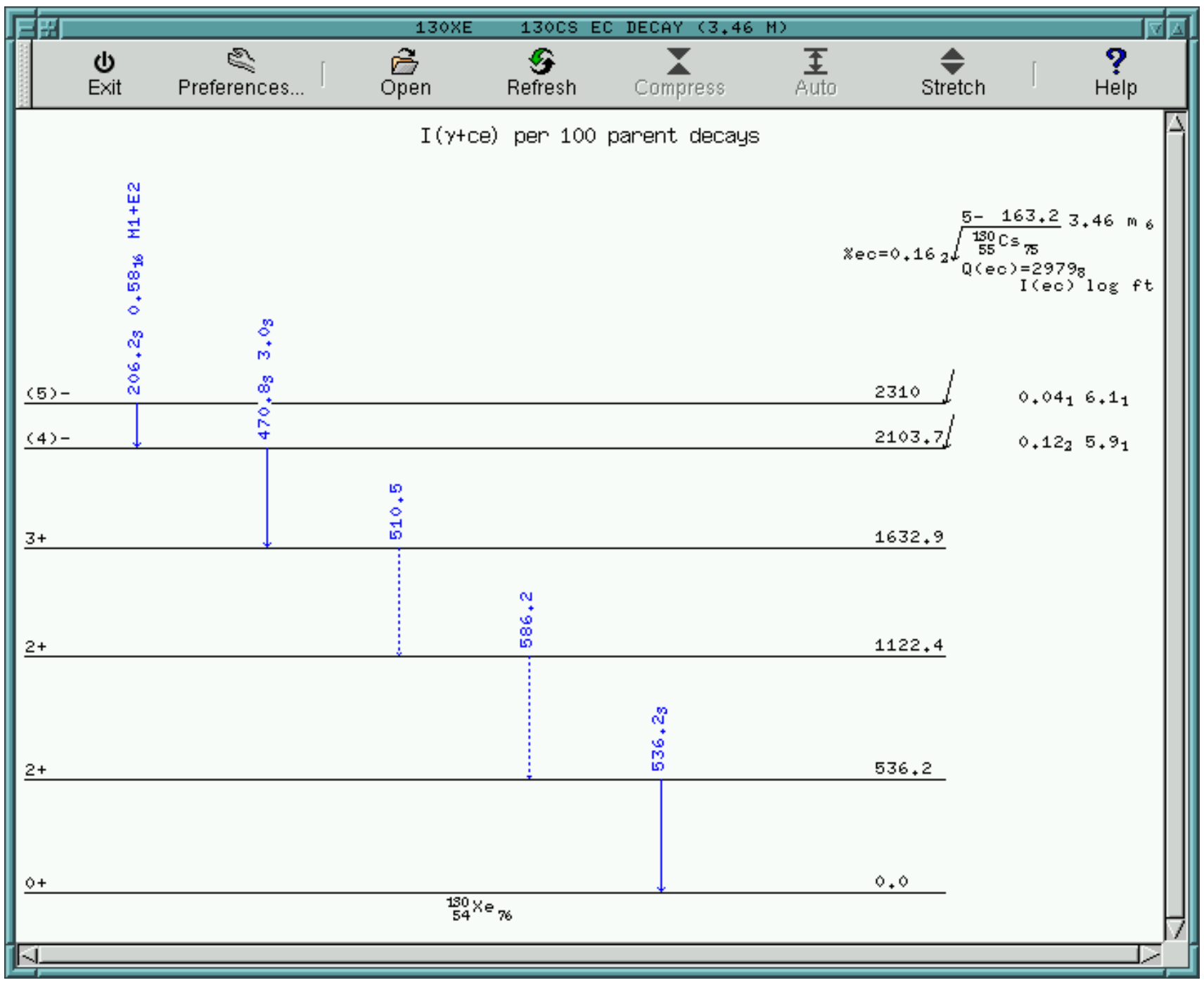

Fig. 8

## Allocation and tuning of the ENSDF_toolbox package

The ENSDF_toolbox is the shell program for a number of other programs. It provides the unified interface for these programs. For the including of the external programs into the package some tuning may be necessary.

Rules for the extension of the package may be found in the file ~/.ENSDF_toolbox/help/{language}/index.html in the section “Additional options”

Two different variants of the installation are possible, namely, system wide (for all users) and local (for the single user) installations. In the first case programs are installed into the directory /usr/local/bin; that directory is described in the variable of environment PATH. In the second case programs are installed into the directory ~/bin in the home user’s directory. If that directory doesn’t exist, it is necessary to create it and to add the description into PATH. For that purpose add to the file ~/.profile next line:

PATH=~/bin:$PATH

The directory .gnome is situated in the user’s home directory. That directory contains configuration files for programs whish are based on the gnome library and underlying libraries. These files have the text format. Configuration file for the ENSDF_toolbox package has the same name as the name of package itself, namely, ENSDF_toolbox. This file contains several sections with headers placed into the square brackets, e.g. [Misc]. Variables for the given section are placed after header and have the format:

*name_of_variable=value*

In this text variables are described as

*name _of_section/name_of_variable*

Some system and service programs are described in these variables, see below. In some cases programs may be listed in a roll, e.g. help browsers Firefox, Mozilla, Opera, etc. In that case two variables are used, one for the roll of programs and other for the ordinal number of the program in the roll. The program may be chosen from the menu “File/Preferences”.

The link of the main shell ENSDF_toolbox with programs which need the terminal dialog is provided by special “subshell” programs. These programs emulate the terminal dialog and transfer of data between the shell and the given program. The example of the “subshell” program for the BrIcc program is presented in the Attachment.

The list of the programs in the ENSDF_toolbox package includes:

1. Archivers and packers, which are used for the reading of the packed files in the ENSDF format. Roll of archivers is determined by the variable Misc/ENSDF_readers in the configuration file ENSDF_toolbox.

2. Parts of the ENSDF_toolbox program package. The program file_selection is used for the selection of the data for the reading and saving. The programs get_sets and data_sets_selection are used for the selection of the datasets; the program rm_cards makes the selection of records in datasets. Programs band_select and band_view are used for the selection of records concerning the rotational bands and for viewing these bands. The EVE program (EValuator’s Editor) is used for the editing of the data in the ENSDF format. Corresponding variables are Misc/ENSDF_editors, Misc/ENSDF_editor, Misc/ENSDF_viewers, Misc/ENSDF_viewer. The program data_set_viewer is used for the viewing of the level schemes; corresponding variables are Misc/ENSDF_scheme_viewers and Misc/ENSDF_scheme_viewer.

3. The Standard BNL programs (http://www.nndc.bnl.gov) for the processing of the ENSDF datasets: ALPHAD, BrIcc, GABS, GTOL, HSICC, LOGFT, RadList, RULER, COMTRANS, ENSDAT, TREND, FMTCHK, PANDORA and corresponding data files for these programs. Data files which are used by the program should be placed in the same directory as the program.

4. Checkers which were developed in PNPI for the work on the correction of the ENSDF database: chk_ENSDF, check_CROSS, check_FL, check_ref, chk_brackets, chk_digit_letter. This roll may be enhanced.

5. Helpers and viewers, such as Firefox, Mozilla etc. for Help files; xterm, gnome-terminal for text files; gimp, gv, xv for PostScript files.

All these programs should be emplaced in the /usr/local/bin directory or in the user’s ~/bin directory, depending on the variant of the installation, or described in the variable of environment PATH.

The program package ENSDF_toolbox is free distributed under GNU general public license as the non-commercial product. The presented release has the status of beta-version, distributed “as is” and is intended for testing. All remarks, proposals, additions will be accepted with acknowledgement. The program package, instruction for the installation and usage may be downloaded from the developer's site http://georg.pnpi.spb.ru.

The work is supported by Russian Foundation for Basic Research, project № 09-07-00387-a.

# Attachment

```
/*****************************************************************
*                                                                 *
*                              exec_BRICC.c                       *
*                                                                 *
*    Subshell program for BrIcc.                                  *
*                                                                 *
*    Author Georgy Shulyak (georg@georg.pnpi.spb.ru).             *
*                                                                 *
*    History.                                                     *
* 1. 19-Sep-2007 - start (with ideas from ext_BRICC.c).           *
*                                                                 *
*****************************************************************/

/* All the #include directives are common for all the interface */
/* progs.*/
#include <errno.h>              /* for errno */
#include <popt.h>               /* for struct poptOption, ... */
#include <stdlib.h>             /* for NULL, EXIT_..., ... */
#include <string.h>             /* for strcpy(), strcat(), ... */
#include <sysexits.h>           /* for EX_UNAVAILABLE */
#include <unistd.h>             /* for _exit() */
#include "exec_daughter.h"      /* for exec_daughter() */
#include "exec_program_dialog.h"/* for dialog_out_string() */
#include "include/gettext.h"    /* for localization */
#include "include/search_file_by_path.h"

#define OWNNAME  "exec_BRICC" /* name of subshell prog. */
#define PROGNAME "bricc" /* name of exe prog. */
#define ENV      "BrIccHome=" /* name of variable of envir. */

/* Next variables are set up by options at start of subshell */
/* and used in the dialog with BrIcc: */
static char *input_name            = NULL;
static char *listing_name          = NULL;
static char *output_name           = NULL;
static char *report_name           = NULL;
static char *new_G_SG_name         = NULL;
static char *comparision_G_SG_name  = NULL;
static int   CC_for_all_subshells   = 0;
static int   CC_for_all_transitions = 0;

static void dialog(     /* dialog*/
    FILE *out_file      /* stream for output (in dialog)  */
    )                   /*   value of function:           */
    )                   /* absent.                  */
{
/* emulation of the input from the keyboard */
    /* Output Files:" */
    /* Complete calculations report, (Def: BrIcc.lst): */
```

```
    dialog_out_string(
        out_file,"%s\n",report_name);
    /* New G/SG records, (Def: Cards.new): */
    dialog_out_string(
        out_file,"%s\n",new_G_SG_name);
    /* G/SG (New/Old) comparision report, (Def: Compar.lst):*/
    dialog_out_string(
        out_file,"%s\n",comparision_G_SG_name);
    /* Execution control: */
    /* List conversion coefficients for all subshells (Def. N): */
    dialog_out_string(
        out_file,"%s\n",CC_for_all_subshells   ? "Y" : "N");
    /* Calculate conversion coefficients for all transitions */
    /* (Def. N): */
    dialog_out_string(
        out_file,"%s\n",CC_for_all_transitions ? "Y" : "N");
}
/* options */
static struct poptOption opt[] = {
    { "input-file",             'i', POPT_ARG_STRING,
       &input_name,             0,
       N_("use input file"),                   "FILENAME" },
    { "output-file",            'o', POPT_ARG_STRING,
       &output_name,            0,
       N_("use output file"),                  "FILENAME" },
    { "report-file",            'r', POPT_ARG_STRING,
       &report_name,            0,
       N_("report to file"),                   "FILENAME" },
    { "new-G-SG",               'n', POPT_ARG_STRING,
       &new_G_SG_name,          0,
       N_("new G/SG file"),                    "FILENAME" },
    { "comparision-G-SG",       'c', POPT_ARG_STRING,
       &comparision_G_SG_name,  0,
       N_("comparision G/SG file"),            "FILENAME" },
    { "listing-file",           'l', POPT_ARG_STRING,
       &listing_name,           0,
       N_("listing to file"),                  "FILENAME" },
    { "CC-for-all-subshells",   'S', POPT_ARG_NONE,
       &CC_for_all_subshells,   0,
       N_("calculate CC for all subshells"),   NULL       },
    { "CC-for-all-transitions", 'T', POPT_ARG_NONE,
       &CC_for_all_transitions, 0,
       N_("calculate CC for all transitions"), NULL       },
      POPT_AUTOHELP
      POPT_TABLEEND
};
int main(           /* dialog          */
    int  argc,      /* number of arguments at start of program */
    char **argv     /* list of arguments at start */
    )
{
```

```
/* common for all subshells */
    poptContext popt_context;
    int         retval;
    char      **other_args;
    char       *daughter_name = PROGNAME;
#define DAUGHTER_NAME_INDEX 0
#define INPUT_NAME_INDEX    1
    char       *daughter_argv[] = { NULL, NULL, NULL };
    char       *env = NULL;
/* common for all the subchells: */
    popt_context =
        poptGetContext(NULL,argc,(const char **)argv,opt,0);
    while ((retval = poptGetNextOpt(popt_context)) >= 0) { };
    if (retval < -1) {
        fprintf(
            stderr,"%s: %s\n",
            poptBadOption(popt_context,POPT_BADOPTION_NOALIAS),
            poptStrerror(retval));
        poptFreeContext(popt_context);
        fclose(stdout);
        return(EXIT_FAILURE);
    };
    other_args = (char **)poptGetArgs(popt_context);
    if (other_args != NULL && other_args[0] != NULL)
        daughter_name = other_args[0];
    retval = EXIT_SUCCESS;
/*setup of variable of environment for BrIcc: */
    {
        char *path_env;
        char *full_path = NULL;
        char *ptr;
        /* setup of variable of environment BrIccHome. */
        /* working files in the same place as */
        /* exe file. */
        if ((path_env = getenv("PATH")) == NULL) {
            fprintf(stderr,_("No PATH environment\n"));
            _exit(EXIT_FAILURE);
        };
        if ((
            full_path =
                search_file_by_path(path_env,daughter_name,0)
        ) == NULL) {
            fprintf(
                stderr,
                _("Executable not found '%s'\n"),
                daughter_name);
            _exit(EX_UNAVAILABLE);
        };
        if ((ptr = strrchr(full_path,'/')) != NULL) *ptr = '\0';
        if ((
            env = malloc(sizeof(ENV)-1+strlen(full_path)+1)
```

```
        ) == NULL) {
            fprintf(stderr,_("Unsufficient memory\n"));
            fflush(stderr);
            free(full_path);
            _exit(EXIT_FAILURE);
        };
        strcpy(env,ENV);  strcat(env,full_path);
        if (
            putenv(env)
        ) {
            fprintf(
                stderr,
                _("exec_BRICC(): putenv('%s') error. errno=%d\n"),
                env,errno);
            fflush(stderr);
            free(full_path);
            free(env);
            _exit(EXIT_FAILURE);
        };
        free(full_path);
    };
/* input filename for BrIcc: */
    daughter_argv[DAUGHTER_NAME_INDEX] = daughter_name;
    daughter_argv[INPUT_NAME_INDEX]    = input_name;
/* common for all the subshells: */
    retval =
        exec_daughter(
            OWNNAME,
            daughter_name,
            daughter_argv,
            listing_name,NULL,dialog,NULL);
    poptFreeContext(popt_context);
    return(retval);
}
```

## References

1. J.K. Tuli. Evaluated Nuclear Structure Data File, A Manual for Preparation of Data Sets, BNL-NCS-51655-01/02-Rev Formal Report (February 2001).

2. M.A. Preston. The Theory of Alpha-Radioactivity, Phys. Rev. 71, 865 (1947).

3. R.S. Hager, E.C. Seltzer. Internal conversion tables, Nucl. Data A4, 1 (1968); R.S. Hager, E.C. Seltzer, Nucl. Data Tables, A6, 1, (1969).

4. I.M. Band, M.B. Trzhaskovskaya, C.W. Nestor, Jr., P.O. Tikkanen and S. Raman. Dirac-Fock internal conversion coefficients, At. Data Nucl. Data Tables 81, 1 (2002).

5. N.B. Gove and M.J. Martin. Log-ft Tables for Beta Decay, Nucl. Data Tables A10, 206 (1971).

6. T.W. Burrows. ENSDF Analysis and Utility Codes, INDC(NDS)-452, part 2, 105, (2004).

7. G. I. Shulyak, A. A. Rodionov. Programs for the work with ENSDF format files: Evaluator's editor EVE, Viewer for the nuclear level schemes; Report PNPI 2821 2009, 15 p.